\documentclass[conference]{IEEEtran}
\IEEEoverridecommandlockouts
\usepackage[letterpaper, left=0.59in, right=0.59in, bottom=0.59in, top=0.50in]{geometry}

\usepackage{cite}
\usepackage{amsmath,amssymb,amsfonts}
\usepackage{algorithmic}
\usepackage{graphicx}
\usepackage{textcomp}
\usepackage{xcolor}
\usepackage{tabularx}
\usepackage{multirow}

\def\BibTeX{{\rm B\kern-.05em{\sc i\kern-.025em b}\kern-.08em
    T\kern-.1667em\lower.7ex\hbox{E}\kern-.125emX}}
\begin{document}

\title{Correlating Power Outage Spread with Infrastructure Interdependencies During Hurricanes\\
\thanks{Notice: This manuscript has been authored by UT-Battelle, LLC under
Contract No. DE-AC05-00OR22725 with the U.S. Department of Energy.
The publisher, by accepting the article for publication, acknowledges that the
U.S. Government retains a non-exclusive, paid-up, irrevocable, worldwide
license to publish or reproduce the published form of the manuscript, or
allow others to do so, for U.S. Government purposes. The DOE will provide
public access to these results in accordance with the DOE Public Access Plan
http://energy.gov/downloads/doe-public-access-plan).}
}

% \author{\IEEEauthorblockN{Avishek Bose\IEEEauthorrefmark{1}, Sangkeun Lee\IEEEauthorrefmark{1}, Narayan Bhusal\IEEEauthorrefmark{2}, Supriya Chinthavali\IEEEauthorrefmark{3}}
% \IEEEauthorblockA{\IEEEauthorrefmark{1}Computer Science and Mathematics Division, Oak Ridge National Laboratory, \textit{\{bosea, lees4\}@ornl.gov}}
% \IEEEauthorblockA{\IEEEauthorrefmark{2}, Electrification and Energy Infrastructures Division, Oak Ridge National Laboratory, \textit{bhusaln@ornl.gov}}
% \IEEEauthorblockA{\IEEEauthorrefmark{3}Geospatial Science and Human Security Division, Oak Ridge National Laboratory, \textit{chinthavalis@ornl.gov}}
% }

\author{\IEEEauthorblockN{Avishek Bose, Sangkeun Lee, Narayan Bhusal, Supriya Chinthavali}
\IEEEauthorblockA{Oak Ridge National Laboratory (ORNL), \textit{\{bosea, lees4, bhusaln, chinthavalis\}@ornl.gov}}
}

\IEEEaftertitletext{\vspace{-1\baselineskip}}

\maketitle

\begin{abstract}
Power outages caused by extreme weather events, such as hurricanes, can significantly disrupt essential services and delay recovery efforts, underscoring the importance of enhancing our infrastructure's resilience. This study investigates the spread of power outages during hurricanes by analyzing the correlation between the network of critical infrastructure and outage propagation. We leveraged datasets from Hurricanemapping.com, the North American Energy Resilience Model Interdependency Analysis (NAERM-IA), and historical power outage data from the Oak Ridge National Laboratory (ORNL)'s EAGLE-I system. Our analysis reveals a consistent positive correlation between the extent of critical infrastructure components accessible within a certain number of steps (k-hop distance) from initial impact areas and the occurrence of power outages in broader regions. This insight suggests that understanding the interconnectedness among critical infrastructure elements is key to identifying areas indirectly affected by extreme weather events.

\end{abstract}


\vspace{-3mm}

\section{Introduction}

Extreme weather events, such as hurricanes, can cause power outages with far-reaching consequences, highlighting the need for a deeper understanding of how these outages spread to strengthen our infrastructure's resilience. It's crucial for planners, emergency responders, and experts to accurately evaluate the potential impacts of such events and pinpoint the vulnerable hotspots.

However, assessing the extent of power outages and their impacts is a complex task. This complexity arises because power outages can occur unpredictably, affecting areas without obvious connections to or even far from the main zones of the weather event. Thus, it's essential to study the network of critical infrastructure and its interdependencies to understand how disruptions to key components can lead to widespread effects across various regions.

In this study, our analysis utilized datasets from three distinct sources. The first dataset, from hurricanemapping.com~\cite{HurricaneMapping2023}, provides forecasted wind swath data from historical hurricanes, offering a broad view of different wind speed ranges. The second dataset, from the NAERM-IA, includes a graph depicting the interdependencies among various critical infrastructure elements in the U.S., such as electric buses and transmission lines. This dataset helps us identify energy components affected by hurricanes and their connections to other infrastructure elements. The final dataset, historical power outage data from ORNL's EAGLE-I system, enables us to determine the number of affected customers and the duration of outages in the areas hit by hurricanes. By leveraging these datasets together, we conducted a correlation analysis to understand the relationship between the infrastructure network's structure and the spread of power outages during extreme weather events.

For the historical hurricanes analyzed, our research showed a positive correlation (greater than 0.6) between the extent of critical infrastructure components accessible within a k-hop distance from the starting nodes in each county and the incidence of power outages affecting customers. This result indicates that leveraging network analysis, which considers the interdependencies among critical infrastructure elements, can effectively identify areas without power impacted indirectly by extreme weather events.

%$The structure of this paper is organized as follows: Section 2 introduces the datasets and methods utilized in our study. Section 3 summarizes our results and findings. Finally, Section 4 offers concluding remarks and outlines future directions.
\vspace{-2mm}
\section{Data and Methods}

\subsection{Hurricane Data}
Hurricanemapping.com provides data from extreme weather events, such as hurricanes and tropical storms. Specifically, we used hurricane forecast wind swath data. Each forecast advisory generates a polygon map file in shapefile format, which forecasts wind swaths for the upcoming 72 hours. These forecasts include three levels of wind speeds: 39 miles per hour (mph), 58 mph, and 74 mph, represented by tiers of polygons. In our research, we concentrated on advisories for two notable hurricanes: Ida and Ian, specifically choosing advisory number 14 and 18 for Ida and number 20 and 27 for Ian.  Hurricane Ida impacted counties in Louisiana, Alabama, Mississippi, and Arkansas and for Hurricane Ian, Florida, Georgia, South Carolina, North Carolina, and Virginia were under the impact. From the advisories for these hurricanes, we identified the regions of interest and the relevant time frames for our study.

\subsection{Historical Power Outage Data}
% Since 2014, the (ORNL)'s EAGLE-{$I^{TM}$}, an interactive system designed to enhance situational awareness by monitoring the U.S. energy infrastructure, has been gathering data on power outages across the United States\cite{osti_1975202}. The data captured by EAGLE-I includes records of the total number of customers experiencing power outages in specific geographical areas at the county level, alongside data on the utility companies involved, with updates provided at 15-minute intervals.
This work leverages the publicly available power outage data for the United States Obtained from EAGLE-I \cite{osti_1975202}. The EAGLE-I data records the total number of customers experiencing power outages in specific geographical areas at the county level, alongside data on the utility companies involved, with updates provided at 15-minute intervals. For our analysis, we extracted data on the number of affected customers per county during the 72 hours covered by the hurricane wind swath data. Although EAGLE-I's outage data is originally in a time-series format with 15-minute granularity, we focused on determining the maximum number of customers impacted during this period for each county. This maximum figure is significant because it offers a snapshot of the peak impact within the timeframe, providing a clear indicator of the hurricane's effect on local power distribution.

\subsection{NAERM-IA Network Data}
The North American Energy Resilience Model (NAERM) is a U.S. Department of Energy (DOE) initiative designed to enhance the reliability and resilience of energy delivery across multiple sectors. A key component of this initiative, developed by ORNL, is the Interdependency Analysis (IA). This tool is designed to assess the cascading effects of damage within the power system across an interconnected network. For our study, we utilized the back-end graph data from NAERM-IA. This comprehensive graph incorporates 31 different types of critical infrastructure components, including energy-related elements (such as generators, transmission line branches, and load buses) and non-energy-related elements (such as fire stations, police stations, and banks). In the graph, nodes represent these components, and edges denote the potential interdependencies between them. For example, an edge connecting two transmission lines indicates that a disruption in one could affect the other. Likewise, an edge from a load bus to a police station illustrates that a power disruption could lead to outages at the police station. 

Each vertex in the graph is assigned geographical coordinates. By selecting wind swath data corresponding to a hurricane event, we can identify energy-related critical infrastructure components within the hurricane's impact zone. Following this identification, we conduct a k-hop neighborhood search with k values from 1 to 5 to determine components that are directly or indirectly dependent on the network. These identified nodes are then considered to be potentially impacted by the hurricane. Subsequently, we tally the number of impacted components within each county and correlate these figures with the maximum number of customer outages recorded during the hurricane's impact, obtained from the EAGLE-I data. A higher correlation value indicates a significant relationship between the topology of the interdependency network and the actual spread of power outages, highlighting the value of interdependency analysis in predicting the effects of such disruptive events on critical infrastructure.

\vspace{-1mm}
\section{Results}
\begin{figure}[htbp]
\centering
% \centerline{\includegraphics{ida_14.png}}
\includegraphics[width=0.5\textwidth, height=0.28\textheight]{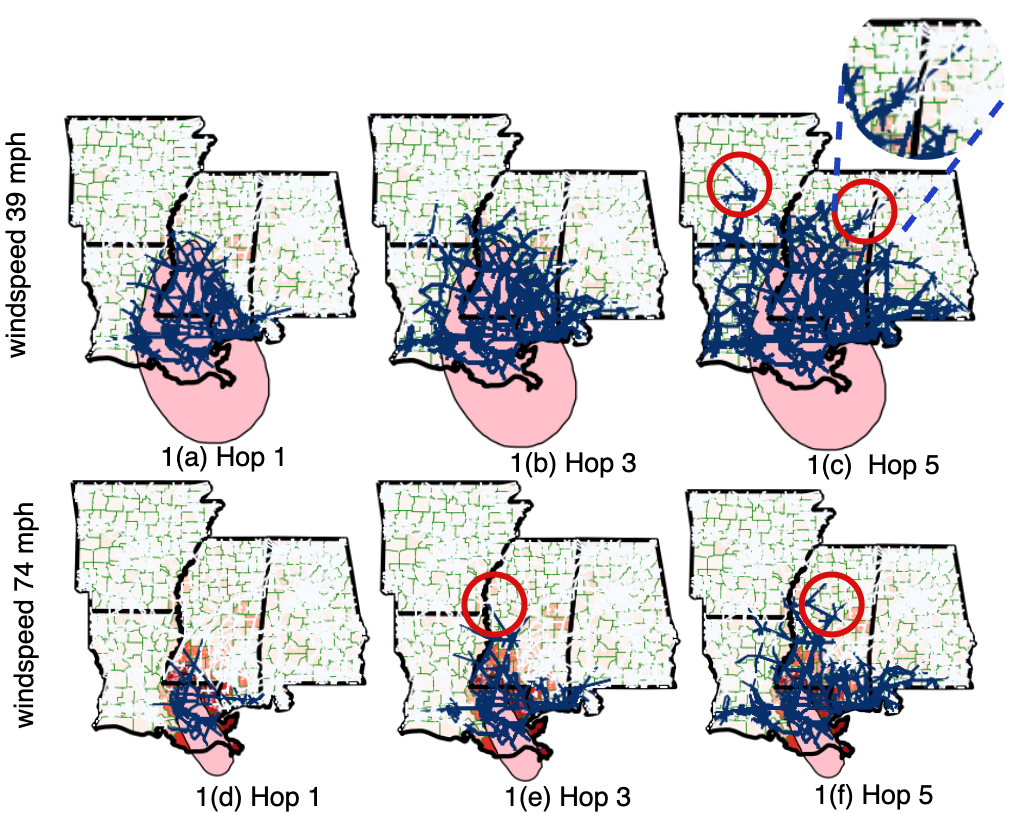}
\caption{Power outage map (next 72 hours) of affected counties with impacted infrastructures from NAERM-IA tool for hop-1, hop-2, and hop-3, for two wind speed levels of IDA 14 advisory}
\label{ida_14}
\end{figure}

Table~\ref{tab1}  presents the Pearson Correlation Coefficients for different configurations in the context of Hurricanes Ida and Ian. Notably, correlations consistently exceed 0.5 for IDA-14 and IAN-20 advisories (extreme weather conditions), indicating a positive relationship. A detailed analysis from Hop-1 to Hop-5 shows stable correlations, albeit with minor fluctuations. Remarkably, correlations for k values greater than 1 occasionally surpass those for direct connections, highlighting the significance of indirectly connected components in understanding the full extent of power outage propagation. Furthermore, an increasing trend in correlation values with more intense wind speeds at higher k values suggests that stronger winds amplify indirect impacts within the network. It is also evident from the rows for advisory number 18 of IDA and 27 of IAN that when wind speed becomes low the correlation values decrease but remain almost stable across hop numbers (indirectly connected components) indicating NAERM-IA's capability to represent power outage propagation. Figure \ref{ida_14} (due to the page limitation only two wind speed forecasts 39 mph, and 74 mph areas are displayed for hop1, hop3, and hop5) illustrates the power outage spread during IDA-14 advisory with IA tool neighborhood search up to 5 hops from the wind speed forecast area. Areas seemingly remote and undisturbed are marked by red circles (one zoomed in subfigure 1(c) ) likely to be impacted very soon. This finding is pivotal, as it reveals that high wind speeds inflict not only immediate damage but also trigger broader, more intricate cascading effects across the interlinked power infrastructure.

\vspace*{-4mm}
\begin{table}[htbp]
\caption{Correlation Analysis Result}
\begin{center}
\begin{tabular}{|m{1.25cm}|m{0.8cm}|c|c|c|c|c|c|}
\hline
\multicolumn{7}{|m{8cm}|}{\textbf{Hurricane IDA 2021 (Eagle-I outage data selected for a time period ranging from 2021-08-28 to 2021-09-05) }}\\
\hline
\textbf{Hurricane Advisory} & \textbf{Force Wind Speed}& \textbf{Hop-1}& \textbf{Hop-2}& \textbf{Hop-3}& \textbf{Hop-4}& \textbf{Hop-5} \\
\hline
\multirow{3}{4em}{IDA-14} & 39mph &0.6077 & 0.6022 & 0.6091 & 0.6185 & 0.6221\\
& 58mph& 0.7773 & 0.7696 & 0.7677 & 0.7761 & 0.7870\\
& 74mph& 0.7708 & 0.7497 & 0.7463 & 0.7597 & 0.7776 \\
\hline
IDA-18 & 39mph & 0.2749 & 0.2745 & 0.2741 & 0.2804 & 0.2896\\
\hline
\multirow{3}{4em}{IAN-20} & 39mph & 0.5763 & 0.5580 & 0.5480 & 0.5473 & 0.5531\\
& 58mph& 0.7412 & 0.7254 & 0.7159 & 0.7130 & 0.7106\\
& 74mph& 0.7180 & 0.7199 & 0.7306 & 0.7421 & 0.7537\\
\hline
\multirow{2}{4em}{IAN-27} & 39mph & 0.5421 & 0.5146 & 0.4970 & 0.4846 & 0.4820\\
& 58mph& 0.4593 & 0.4535 & 0.4584 & 0.4665 & 0.4779\\
\hline
\end{tabular}
\vspace*{-5mm}
\label{tab1}
\end{center}
\end{table}

% \begin{figure}[htbp]
% \centering
% % \centerline{\includegraphics{ida_13.png}}
% \includegraphics[width=0.5\textwidth, height=0.55\textheight]{ida_13.png}
% \caption{IDA 13}
% \label{fig}
% \end{figure}

\vspace{-5mm}

% \begin{figure}[htbp]
% \centering
% % \centerline{\includegraphics{ida_14_2.png}}
% \includegraphics[width=0.5\textwidth, height=0.45\textheight]{ida_14_2.png}
% \caption{IDA 14}
% \label{fig}
% \end{figure}
\vspace{-3mm}
\section{Conclusion and Future Work}
In conclusion, these results emphasize the necessity of a holistic approach that includes directly and indirectly affected areas when analyzing the impact of power outages caused by extreme weather events. For future work, we plan to expand our correlation analysis to include a wider range of hurricane events and advisories and will implement various normalization techniques to address any potential biases in our findings. We also plan to investigate the reasons for obtaining high and low correlation values in different cases when using the NAERM-IA tool, and whether and to what extent the variations in the network structure of the IA tool affect these correlation values.

\bibliographystyle{IEEEtran} % Specifies the style of the bibliography
% \bibliography{ref} % The filename of your .bib file without the extension

\end{document}